# A Novel Approach for Mining Similarity Profiled Temporal Association Patterns

**Vangipuram Radhakrishna[1], Dr.P.V.Kumar[2], Dr.V.Janaki[3]**

[1] Faculty of Information Technology, VNR Vignana Jyothi Institute of Engineering and Technology, Hyderabad, 500090, India
[2]Professor, Department of Computer Science and Engineering, Osmania University, Hyderabad, 500090, India
[3]Professor & Head, Department of Computer Science and Engineering, Vaagdevi Engineering College, Warangal, 506001, India.
[1]Email :vrkrishna2014@gmail.com

**Abstract.** The problem of frequent pattern mining from non-temporal databases is studied extensively by various researchers working in areas of data mining, temporal databases and information retrieval. However, Conventional frequent pattern algorithms are not suitable to find similar temporal association patterns from temporal databases. A Temporal database is a database which can store past, present and future information. The objective of this research is to come up with a novel approach so as to find similar temporal association patterns w.r.t user specified threshold and a given reference support time sequence using concept of Venn diagrams. For this, we maintain two types of supports called positive support and negative support values to find similar temporal association patterns of user interest. The main advantage of our method is that, it performs only a single scan of temporal database to find temporal association patterns similar to specified reference support sequence. This single database scan approach hence eliminates the huge overhead incurred when the database is scanned multiple times. The present approach also eliminates the need to compute and maintain true support values of all the subsets of temporal patterns of previous stages when computing temporal patterns of next stage.

## 1. INTRODUCTION

In last four decades, significant research contribution from database community is towards studying temporal databases and various aspects of temporal information systems. In 1986, a summary of temporal database research discussed at various symposiums, workshops and work carried out at various research labs was first published in ACM SIGMOD Record. The importance of temporal databases coined and came in to existence with IEEE Data Engineering devoting a complete issue for temporal databases in 1988. Consequently in year 1990 and 1992, two research papers that contributed to survey on temporal databases were published. In (Gultekin Ozsoyoilu et.al, 1995), authors perform survey on temporal and real time databases.

A detailed literature survey on temporal databases, temporal database methodologies and techniques, applications is discussed in (Srivatsan Laxman et.al, 2006). The theory of temporal databases, design and implementation is discussed in very interesting manner by (Tansel et.al, 1993) which is edited volume of papers presented at symposiums and workshops as various chapters. In (Matteo Golfarelli et.al, 2009), authors perform detailed literature survey on temporal data warehouses. (Alexander.et.al, 2003) discussed database support for database applications. (Tansel et.al, 2007) work involves finding temporal association rules from temporal databases. An approach to discover temporal association rules from publication databases is addressed in research contributed by authors (Chang-Hung Lee et.al, 2001). In (Toon Calders et.al, 2001), authors define the approach for finding frequent items using the approximization method by computing upper bound and lower bound of items. An approach for finding frequent items in spatio temporal databases is discussed in (Cheqing Jin.et.al, 2004). A method for finding temporal frequent patterns using TFP-tree is discussed in (Long Jin et.al, 2006).

The authors (Jin Soung Yoo et.al 2009), use the Euclidean distance measure as dissimilarity measure to find temporal association patterns of user interest. This work is further extended in (Jin Soung Yoo et.al, 2012**).** The authors approach includes finding upper-lower bound, lower-lower bound and lower bound distances to discover similar temporal patterns. They define two methods for discovering temporal patterns. The drawback of this approach is that the algorithm designed requires knowing true support values of temporal pattern of size, k, for deciding if a pattern of size, k+1 is temporally similar or temporally dissimilar. This initiates the necessity for scanning database more than once to obtain true support sequences of temporal pattern at level-k. In present approach of finding temporal association patterns, we overcome this dis-advantage and also eliminate the necessity to compute true support values of its sub patterns and the need to consider support sequence, support values of all possible subsets of size-k temporal patterns for finding lower bound and upper bound support sequences of temporal patterns of size, k+1. A recent survey on temporal databases and data mining techniques is discussed in (Vangipuram et.al, 2015).

## 2. MOTIVATION

This research is mainly motivated from research contribution of authors (Jin Soung Yoo et.al 2012). The present approach which we propose is the novel approach which can be used to efficiently find similar temporal association patterns of interest w.r.t a given reference support time sequence, R and a user specified threshold constraint to overcome following disadvantages of (Jin Soung Yoo et.al 2012).

1. We eliminate, multiple scan of temporal database required to find support sequences of temporal patterns generated at different levels.
2. We overcome, the disadvantage of approach followed in (Jin Soung Yoo et.al 2012), which requires repeated scanning of the temporal database to know true support values of temporal association patterns of level-(k-1) when finding temporal association patterns at level-k.
3. We overcome the requirement to know the true support sequence values of all subsets of an itemset as followed in (Jin Soung Yoo et.al 2009).

The problem of mining frequent patterns in databases is addressed extensively in the literature. Conventional algorithms designed to obtain frequent patterns are not applicable to find frequent patterns from temporal database of time stamped transaction sets. This is because, all these algorithms which are used to find frequent patterns are essentially based on support values which are actually independent of time stamp of transactions.

In our case i.e. for temporal database of time stamped transactions, these transactions have time stamps and hence the support of the itemsets is in the form of a vector representing support values computed for each time slot. This makes the conventional approach unsuitable to obtain frequent temporal patterns. Also, the popular Euclidean distance measure which is used to find distance between any two vectors does not satisfy the monotonicity property (Jin Soung Yoo et.al 2012). In the present work, we consider the problem of mining similarity profiled temporal patterns from set of time stamped transaction sets of a given temporal database. We demonstrate using a case study how the proposed approach may be used to find similar temporal frequent patterns.

## 3. MOTIVATION
### 3.1. Problem Definition

To apply concept of Venn diagrams and to introduce two types of support values called positive and negative support values so as to find similar temporal association patterns of user interest from the temporal database of time stamped transaction sets, performing only a single database scan. This is achieved through designing and validating the suitability of proposed approach using concept of Venn diagrams to find temporal association patterns. The proposed approach of finding temporal patterns is based on algorithm design strategy called dynamic programming technique as it makes use of support time sequence of temporal patterns obtained in previous stage aiming at both time and space optimizations.

### 3.2. Problem Statement

Given a

i)  A Finite set of singleton items denoted by I ,

ii)  $D_{temporal}$ , formally represented as $D_{temporal} = D_1 \cup D_2$ .......$\cup D_n$ is a temporal database of time stamped transaction sets such that for any value of i and j, whenever $i \neq j$ , we have both $D_i$ and $D_j$ to be disjoint.

iii)  $T = t_1 \cup t_2$ .......$\cup t_n$ , is the time period such that for any two values i, j whenever $i \neq j$ , we have both $T_i$ and $T_j$ to be disjoint sets.

iv)  Each transaction record, d in $D_{temporal}$, is formally represented as a 2-tuple <timestamp, itemset>, where timestamp and itemset, $I$ are subsets of T and I respectively.

v)  $R = < r_1 , r_2 , r_{3, .....} r_n >$ is a reference support sequence over time slots $t_1 , t_2$ ......., $t_n$ and a user specified dissimilarity value denoted by $\theta_{th}$

vi)  A distance measure, function of two support sequences P and Q denoted as $f_{similarity} \ (P, Q) : \rightarrow R^n$.

The objective is to "Discover set of all temporal patterns, $I$ or $I_{temporal}$ , which are subsets of I such that each of these itemsets represented by $I$ , satisfy the condition $f_{similarity}$ ( $S_I$, Reference) $\leq \theta_{th}$ where $S_I$ is the sequence of support values of $I$ at time slots $t_1 , t_2$ ......., $t_n$

### 3.3. Objective

The main objective of the present research is to essentially find similar temporal association patterns or frequent patterns w.r.t a specified reference sequence which satisfy the condition $f_{similarity}$ ( $S_I$, Reference) $\leq \theta_{th}$ by scanning the temporal database only once. The basic idea is to make use of Venn diagram concept and effectively use the support sequences of itemsets found initially after the first scan of the database. In order, to find the upper-lower bound and lower-lower bound, lower bound distances we adopt the method outlined in (Jin Soung Yoo et.al 2009). For dissimilarity computation, we use the Euclidean distance measure to find the distance between itemset support sequence and the reference sequence. Since we use the distance measure (also called as similarity measure) to find the temporal association patterns, we call these temporal patterns as similarity profiled temporal association patterns.

### 3.4. Basic Terms and Terminology

In this section we define the basic terms and terminology used for discovering temporal association patterns.

**Positive itemset:** An itemset, $I' \subseteq I$, is called a positive itemset, if we compute the probability of its existence in the temporal database. For example, the itemsets represented as $A$, $AB$ are called positive itemsets. We use terms itemset and pattern interchangeably in this paper.

**Negative Itemset:** An itemset denoted by, $I' \subseteq I$, said to be a negative itemset, if we compute the probability of its absence in the temporal database. For example, the item sets represented as $\bar{A}$, $\overline{AB}$, $\bar{B}$ are called negative itemsets. We may also denote, $\bar{A}$, $\overline{AB}$, $\bar{B}$ as A', (AB)'and B' respectively.

**Positive Support:** The support value computed for positive itemset is called as positive support.

**Negative Support:** The support value computed for negative itemset is called as negative support.

**Support Sequence:** The support sequence of a temporal pattern is an n-tuple denoted by $S_\theta (I') = < S_{t_1}, S_{t_2}, S_{t_3}, ........S_{t_n} >$, where each $S_{t_i}$ is the support value of itemset, $I'$ at time slot, $t_i$. Formally, $S_\theta (I') = U_n \{ S_{ti} \mid$ time slot, $t_i$, varies from $t_1$ to $t_n \}$ where the symbol, $U_n$ ,denotes union of all support values computed for each time slot, $t_i$.

**Positive Support Sequence, $S_{\theta_p}(I')$ :** The support time sequence, $S_\Theta$, of a temporal pattern $I' \subseteq I$ denoted by, $S_\theta (I')$, is defined as positive support sequence, denoted by $S_{\theta_p}(I')$, if and only if, the support of each element, $S_{t_i}$, in the support sequence is obtained for the existence of itemset $I'$ in temporal database.

**Negative Support Sequence:** The support time sequence, $S_\Theta$, of a temporal pattern $I' \subseteq I$ denoted by, $S_\theta (I')$, is called negative support sequence , denoted by $S_{\theta_N}(I')$, if and only if, the support of each element, $S_{t_i}$, in the support sequence is computed for the absence of itemset I' in temporal database. The negative support sequences denote the probability of the temporal pattern not existing in the time slots, $t_1, t_2, ........ t_n$.

**True Distance:** It is formally defined as the actual distance between the support sequence vector of a temporal pattern and the reference support sequence vector, obtained using Euclidean distance.

**Upper Lower Bound (ULB):** The maximum possible distance between upper bound support sequence of a temporal pattern and the reference sequence vector is formally defined as upper-lower bound distance (Jin Soung Yoo et.al 2012).

**Lower Lower Bound (LLB):** The minimum possible distance between lower bound support sequence of a temporal pattern and the reference sequence vector is formally defined as lower-lower bound distance (Jin Soung Yoo et.al 2012). This value indicates that the support value at time slot, $t_i$, cannot be less than this value.

**Pruning:** The process of discarding or eliminating all those temporal patterns which do not satisfy user defined constraints is called pruning. A temporal pattern, denoted by $I'$, is considered temporally similar pattern if and only if, every subset $I'$ of I is also temporally similar.

### 3.5. Proposed Algorithm

The algorithm to find similar temporal association patterns w.r.t reference support sequence and user specified threshold is given below

**Input:**

| | |
|---|---|
| L or I | Finite set consisting of single items |
| $D_{temporal}$ | Temporal database of transactions |
| R | Reference Sequence |
| $f_{similarity}$ | Euclidean Distance measure |
| $\theta$ | Threshold |
| k | Size of Itemset |
| $I$ | Finite set of itemsets |
| $I'$ | Itemset, subset of I |
| N | Number of items in Set, L |

**Output:**

Set of all temporal patterns, $I'$ , which are subsets of I , such that each of these temporal itemsets represented by $I'$, satisfy the condition $f_{similarity}$ $(S_I,R) \leq \theta$ where $S_I$ is support sequence of temporal pattern, $I'$ at time slots $t_1$ , $t_2$ ……., $t_n$

**Step1:**

For every time slot, $t_i$ , compute support values for positive and negative singleton temporal patterns of size, $|S|=1$

**Step2:**

From support values for singleton temporal items, found in Step1, find positive and negative support sequences. These positive and negative support sequences are represented by $S_{\theta_P}$, $S_{\theta_N}$ respectively. The temporal patterns, we consider are categorized into 3 types according to size of temporal patterns, denoted by $|S|$. We consider three cases here i.e temporal patterns of size, $|S|=1$; $|S|=2$; and temporal patterns of size, $|S| > 2$

**Step-3:  Size, $|S|=1$, K=1, Level-1 temporal association patterns (X, Y, Z ….)**

Consider each singleton temporal pattern of size, $|S|=1$. Obtain the distance between each singleton temporal pattern and reference support sequence vector. Compare these distances with user defined threshold value. All these singleton temporal patterns, whose distance from reference support sequence is less than user specified threshold value, are considered similar temporal patterns, while those whose distance is greater than user specified threshold value are treated to be dissimilar temporal patterns. Even though, temporal patterns whose distance is greater than user specified threshold value are considered to be dissimilar temporal patterns, however, we consider to retain all such temporal patterns of size , $|S|=1$ , whose approximate upper lower bound value, $ULB_{approx}$  is less than user specified threshold value. This is mainly done to compute the temporal patterns of size, $|S| \geq 2$.

**Step-4:  Size, $|S|=2$, K=2, Level-2 temporal association patterns (XY, XZ, YZ….)**

Set $|S|=|S|+1$. Generate all possible combinations of temporal patterns of size, $|S|=2$, from temporal association patterns of size, $|S|=1$, retained in previous step-3. Now, temporal itemset combinations generated will be of the form $I_i\,I_j$ where $I_i$ must be mapped to level- (K-1) itemsets of length ($|S| = 1$) and $I_j$  indicates, the singleton temporal item of length equal to one, not present in $I_i$. This step mainly involves finding maximum and minimum support sequence of temporal itemsets of size, $|S| = 2$ respectively.

To compute support sequences of temporal itemset of size, $|S|=2$, we may use the expression,  $I_i\,I_j \;=\; I_j - I_j\,\overline{I_i}$ (or)  $I_i\,I_j \;=\; I_i - I_i\,\overline{I_j}$ . Both these expressions give same result. Here, for each temporal pattern of the form  $I_i\,I_j$ , $I_i$ is the temporal itemset of size, $|S|=1$, whose support sequence is already obtained in Step-3. Also, the support sequence for temporal pattern, $I_j$ is computed in step-3, as this pattern is also of size, $|S|=1$. To compute, support time sequence for temporal pattern of the form, $I_i\,I_j$ , we must compute maximum and minimum possible support sequence vectors of temporal pattern of the form  $I_i\,\overline{I_j}$  and then obtain the corresponding minimum and maximum possible support sequences for temporal pattern or itemset of the form , $I_i\,I_j$ with, $|S|=2$.

Now, find the upper-lower bound , lower-lower bound, and lower bound values for temporal pattern, $I_i\,I_j$ . If the lower bound distance value of temporal pattern, lower bound (LB) $\leq \theta$ , then consider it as similar temporal association pattern, otherwise treat all such itemsets as temporally dissimilar. However, if, the approximate upper lower bound distance value of $I_i \, I_j \leq \theta$, then, retain all such itemsets of the form, $I_i \, I_j$, to generate itemset support sequences of size, |S|>2.

**Step-5: |S|>2, Level- K // generalized for all temporal patterns of size, |S| > 2 i.e. |S|=3, 4, 5…..n**

Set |S| = |K|+1. Generate all possible temporal patterns of size, |S|>2, from temporal association patterns of size, |S|=S-1, retained in previous step. Now, temporal itemset combinations generated will be of the form $I_i \, I_j$ where $I_i$ must be mapped to first |S|-1 sequence of items or level- (K-1) itemsets of length equal to (k-1) and $I_j$ indicates, the singleton temporal item of length equal to one, not present in $I_i$.

For, temporal patterns of size, |S|>2, we have a peculiar situation. This is because, when |S|=1, we know true support values of singleton temporal patterns, as these are obtained by scanning the database directly. This finishes first scan. For |S|=2, we do not compute true support sequences, but we obtain the maximum and minimum possible support sequences of temporal patterns using the proposed approach as in step-4.

So, for temporal patterns of size, |S|>2, such as |S|=3, 4, 5….n, we have four cases to be considered as shown in Eq.2, for computing itemset support sequences. Equation.1, gives expression for finding support sequence of temporal patterns of size, |S|>2.

In Equation.1 below, $I_i \, I_j$ is temporal pattern of size, |S|>2

$$I_i \, I_j = \begin{cases} I_{i_{min}} - I_{i_{min}} \otimes \overline{I_j} \\ I_{i_{max}} - I_{i_{max}} \otimes \overline{I_j} \end{cases}$$

(1)

From Eq.1, we have four sub cases generated as shown below using Eq.2

$$I_i \, I_j = \begin{cases} I_{i_{min}} - UBSTS \, (I_{i_{min}} \otimes \overline{I_j}) \\ I_{i_{min}} - LBSTS \, (I_{i_{min}} \otimes \overline{I_j}) \\ I_{i_{max}} - UBSTS \, (I_{i_{max}} \otimes \overline{I_j}) \\ I_{i_{max}} - LBSTS \, (I_{i_{max}} \otimes \overline{I_j}) \end{cases}$$

(2)

From these support sequences, obtain $I_i \, I_{j_{max}}$ and $I_i \, I_{j_{min}}$, the maximum and minimum support sequence of temporal itemsets of size, |S|>2 respectively. These are called maximum support time sequence and minimum support time sequence of itemset, $I_i \, I_j$, of size, |S|>2. Now, find the upper lower bound and lower lower bound, lower bound distance values for the itemset $I_i \, I_j$. If the value of lower bound distance < user threshold, then consider the corresponding temporal pattern as a similar temporal association pattern, otherwise treat all such itemsets as temporally dissimilar. However, if, the approximate upper lower bound distance value of temporal pattern $I_i \, I_j$ is less than user threshold, $\theta$, retain all such itemsets of the form, $I_i \, I_j$, to find itemset support sequences of next level itemsets of size, |S|=|S| +1. Repeat step-5 till size of temporal itemset is equal to number of items in set, denoted by I or until no further temporal patterns are generated.

**Step-6:** Output all similar temporal association patterns w.r.t reference support sequence satisfying user specified constraints.

### 3.6. Algorithm Explanation

**Step1: Compute support values of both the positive and negative singleton temporal patterns of size, k=1 for each time slot, $t_i$.**

This requires computing the probability of existence of each singleton item of size, k=1 in the temporal database. From these values obtained, find probability values for negative items of size, k = 1. We use the probability of itemset and support interchangeably in this paper.

**Step2: Obtain Positive and Negative Support Sequences $(S_{\theta_P}, S_{\theta_N})$ of singleton items from support values of positive and negative temporal patterns or temporal items obtained in step-1.**

Let the finite itemset, I, consists of items A, B and C then, from support values of positive and negative singleton items obtained for every time slot, find corresponding support sequences for items A, B, C. These are denoted by $S_{\theta_P}(A)$, $S_{\theta_P}(B)$ and $S_{\theta_P}(C)$ and are called positive support sequences. Similarly, from support values of all negative temporal patterns $\bar{A}$, $\bar{B}$, $\bar{C}$ obtained, find corresponding support time sequences for negative temporal patterns. These are denoted by $S_{\theta_N}(\bar{A})$, $S_{\theta_N}(\bar{B})$, $S_{\theta_N}(\bar{C})$ and called as negative support sequences.

**Step-3: Find level-1 similarity profiled temporal patterns.**

Compute the true distance, upper-lower bound, lower-lower bound distances between the single item support sequences obtained in step-2 and reference sequence using the method adopted in (Jin Soung Yoo et.al 2012). If the true distance ≤ threshold value, then this item is considered as similar temporal pattern. However if the true distance computed exceeds the threshold, and corresponding upper lower bound (ULB) distance is less than the user specified threshold, then we retain this itemset without killing, so that it may be used to compute temporal association patterns of the next stage. All such item sets which are considered for retaining in the previous stage are not treated similar temporal association patterns but are only retained for purpose of finding temporal patterns of higher levels.

**Step-4: Set k = k+1. Find the level-k similarity profiled temporal association patterns. Here (K=2, 3, 4….n)**

Generate all possible non-empty itemset combinations of size, k= k+1, using itemsets retained at level-k. For these temporal patterns of size equal to k+1, obtain the maximum possible support sequence, i.e. upper bound support sequence and minimum possible support sequence, i.e lower bound support sequence. This is followed by computation of the lower bound distance of the support sequence of itemset w.r.t the reference sequence. If the lower bound distance, satisfies the threshold constraint then the pattern is temporally similar w.r.t reference or similar temporal pattern. Otherwise, it is not treated as the similar temporal association pattern.

The algorithm or procedure to compute the support sequence is given below.

**Algorithm to generate upper bound support time sequence (UBSTS), lower bound support time sequence (LBSTS) , compute lower bound (LB) distance of temporal pattern of the form $I_i I_j$ of size, k ≥ 2**
{

1. Consider the Equations (3) and (4) to find support time sequence of temporal pattern $I_i I_j$ as given below

$$I_i I_j = I_j - I_j \overline{I_i} \qquad (3)$$

(Or)

$$I_i I_j = I_i - I_i \overline{I_j} \qquad (4)$$

For each generated itemset (also called as association pattern) combination of the form $I_i I_j$ which is of size equal to k, $I_i$ must be mapped to the level–(k-1) itemset of length equal to (k-1) and $I_j$ indicates the singleton item of length equal to 1 which is not present in $I_i$ using equation 3 or equation 4.

2. Compute the upper bound support sequence and lower bound support sequence for association pattern of the form $I_i \overline{I_j}$ using the procedure outlined in section 3.7. From these maximum and minimum possible support sequences computed for pattern $I_i \overline{I_j}$, Obtain corresponding minimum ($I_i I_{j_{MIN}}$) and maximum ($I_i I_{j_{MAX}}$) support sequences for the association pattern denoted by $I_i I_j$ using Equations (5) and (6).

$$I_i I_{j_{MIN}} = I_i - I_i \overline{I_{j_{UBSTS}}} \qquad (5)$$

and

$$I_i \, I_{j \, MAX} \; = \; I_i - I_i \, \bar{I}_{j \, LBSTS} \tag{6}$$

Here $I_i \, I_{j \, MAX}$ and $I_i \, I_{j \, MIN}$ denote maximum and minimum support time sequences of temporal pattern, $I_i \, I_j$.

3. Compute the approximate upper-lower bound and approximate lower-lower bound distance values of temporal pattern, denoted by $I_i \, I_j$ using maximum and minimum support sequence vectors generated in step-3 and using procedure outlined in sub sections 3.7.1, 3.7.2, 3.7.3.

4. Finally, obtain the lower bound distance by summing both upper-lower bound and lower-lower bound distances
   i.e. LB (Lower bound) = ULB (Upper lower bound) + LLB (Lower lower bound)
}

**Step-5: Repeat step-4 till itemset size is equal to number of items in set I or till no further itemsets can be generated.**

For generating candidate items for level k+1 we use Equation 1 and Equation 2. At every stage of generating the itemset combinations as in step-4 and step-5, prune all the itemset combinations even if any subset of these item sets is not considered frequent in previous stages.

For example if XY is not frequent, then even if XZ and YZ are frequent, we can say XYZ is not frequent. However on the fly from stage-i to stage – (i + 1), all those itemsets which satisfy upper-lower bound distance may be retained.

### 3.7. Generating UBSTS, LBSTS support time sequences and ULB, LLB and LB distances

Generating itemset support time sequences is crucial to find the similar temporal association patterns. To generate upper bound and lower bound support sequences we follow the procedure outlined in **(Jin Soung Yoo et.al 2012)**. However the computation of support time sequence for a given itemset or pattern is done using different method using the equations 1 to 6. The earlier approach for finding support time sequence of an itemset requires knowing the support values of all its subsets and also requires scanning the database for actual support values at previous stage to find support sequence of temporal association pattern of next stage.

In our proposed approach, computation of support time sequences for itemset combination $I_i \, I_j$ requires computing only support time sequences for itemset denoted by $I_i \, \bar{I}_j$ and $I_i$. This eliminates the need to maintain support of all subsets of itemset of size k . For the first time, we generate the upper bound and lower bound support sequences for the itemset combination $I_i \, \bar{I}_j$ and use these support sequences, to compute support time sequences of $I_i \, I_j$.

### 3.7.1 Computation of Upper Bound and Lower Bound Support Time Sequences
Let

$$S(I_i) = < S_{i1} , S_{i2} , S_{i3} , \ldots \ldots \; S_{im} >$$

$$S(I_j) = < S_{j1} , S_{j2} , S_{j3} , \ldots \ldots \; S_{jm} > \tag{7}$$

be the support time sequences of items $I_i$ and $I_j$.

The upper bound and lower support time sequences of itemset $I_i \, I_j$ are computed using equations (8) and (9)

$$UBSTS(I_i \, I_j) \; = \; < min(S_{i1} , S_{j1}) , min(S_{i2} , S_{j2}) , min(S_{i3} , S_{j3}) \ldots \ldots \ldots \ldots min(S_{im} , S_{jm}) > \tag{8}$$

$$LBSTS(I_i \, I_j) \; = \; < max(S_{i1} + S_{j1} - 1,0) , max(S_{i2} + S_{j2} - 1, 0) , \ldots \ldots \ldots min(S_{im} + S_{jm} - 1, 0) > \tag{9}$$

### 3.7.2 Computation of Upper-Lower Bound distance
Let R=$< r_1 , r_2 , r_3 , \ldots \ldots r_m > \; = \; < r_i \mid i \leftarrow 1$ to m $>$ be a reference sequence and U $= \; = < U_1 , U_2 , U_3 , \ldots \ldots U_m >$ be the upper bound support time sequence. Now, if we assume $R^{Upper}$ and $U^{Lower}$ to be the subsequence of reference and upper bound support time sequences of length, k, such that for all i varying from 1 to k, the condition $R_i > U_i$ holds good, then the upper-lower bound distance value is computed as ULB-distance(R, U) = distance between $R^{Upper}$ and $U^{Lower}$ of length k using Euclidean distance of k-dimensions.

### 3.7.3 Computation of Lower-Lower bound distance

Let R=< $r_1$, $r_2$, $r_3$, ..........$r_m$ > = < $r_i$ | i ←1 to m > be a reference support sequence and L = < $L_1$, $L_2$, $L_3$, ..........$L_m$ > be the lower bound support time sequence. Now, if we assume $R^{lower}$ and $L^{upper}$ to be the subsequence of reference and lower bound support time sequences of length k, such that for all i varying from 1 to k, the condition $R_i < L_i$ holds good, then the lower-lower bound distance value is computed as LLB-distance(R, L) = distance between $R^{lower}$ and $L^{upper}$ of length k using Euclidean distance of k-dimensions.

### 3.7.4 Computation of Lower Bound distance

The lower bound distance is the sum of ULB (upper lower bound) and LLB (lower lower bound) distances. Mathematically, we may define the lower bound value as given by Equation.10

LB (Lower-bound distance) = ULB (Upper Lower Bound) + LLB (Lower Lower Bound)     (10)

### 4. CASE STUDY

The Table.1 denotes sample temporal database defined over time stamped transactions with finite set, I consisting items {X, Y, Z}. The temporal database is partitioned into two groups of transaction sets performed at time slots, $t_1$ and $t_2$ respectively.

Here we assume user specified threshold value as 0.22 and reference support sequence as <0.4, 0.6> as shown in Table.3.Table.2 contains values of true support time sequences for all possible itemset combinations. Since there are 3 items, there are 7 possible combinations as shown in Table.2.

**Step-1:**

Initially, we start by scanning temporal database to find the positive support value of singleton items X, Y, Z and also corresponding negative support of items $\overline{X}$, $\overline{Y}$, $\overline{Z}$ for each time slot. The support values at time slots $t_1$, and $t_2$, is shown in the table 4 for each positive and negative singleton temporal pattern.

Table 1. Sample Temporal Database

| Database, D1 | | Database, D2 | |
|---|---|---|---|
| Time Slot,T1 | Items | Time Slot,T2 | Items |
| 1 | X | 11 | Y,Z |
| 2 | X,Y,Z | 12 | Y |
| 3 | X,Z | 13 | X,Y,Z |
| 4 | X | 14 | X,Y,Z |
| 5 | X,Y,Z | 15 | Z |
| 6 | Z | 16 | X,Y,Z |
| 7 | Z | 17 | X,Z |
| 8 | X,Y,Z | 18 | Z |
| 9 | Z | 19 | Y |
| 10 | Z | 20 | Y,Z |

Table 2. True support sequences

| Combination | Itemset | Itemset Support Time Sequence |
|---|---|---|
| 1 | X | <0.6,0.4> |
| 2 | Y | <0.3,0.7> |
| 3 | Z | <0.8,0.8> |
| 4 | XY | <0.3,0.3> |
| 5 | XZ | <0.4,0.4> |
| 6 | YZ | <0.3,0.5> |
| 7 | XYZ | <0.3,0.3> |

Table3. Reference Sequence

| Reference | Support Time Sequence |
|---|---|
| Ref | <0.4,0.6> |

Table 4. Support values of singleton items at $t_1$, $t_2$

| | Singleton Itemset | Support at $t_1$ | Support at $t_2$ |
|---|---|---|---|
| Positive Singleton item | X | 0.6 | 0.4 |
| | Y | 0.3 | 0.7 |
| | Z | 0.8 | 0.8 |
| Negative singleton item | $\overline{X}$ | 0.4 | 0.6 |
| | $\overline{Y}$ | 0.7 | 0.3 |
| | $\overline{Z}$ | 0.2 | 0.2 |

Table 5. Positive & Negative Support Time Sequences

| Positive Itemset | PSTS , $S_{\theta_P}$ | Negative Itemset | NSTS, $S_{\theta_N}$ |
|---|---|---|---|
| X | <0.6,0.4> | $\overline{X}$ | <0.4,0.6> |
| Y | <0.3,0.7> | $\overline{Y}$ | <0.7,0.3> |
| Z | <0.8,0.8> | $\overline{Z}$ | <0.2,0.2> |

Table 6. ULB and actual distance of itemset w.r.t Ref

| Itemset | PSTS , $S_{\theta_P}$ | ULB-approx | actual distance |
|---------|------------------------|------------|-----------------|
| X | <0.6,0.4> | 0.2✓ | 0.28 ✗ |
| Y | <0.3,0.7> | 0.1✓ | 0.14 ✓ |
| Z | <0.8,0.8> | 0.0✓ | 0.45 ✗ |

Table 7. Lower Bound distance computation of XY

| Itemset | STS | ULB | LLB | LB |
|---------|-----|-----|-----|-----|
| $XY_{max}$ | <0.3,0.4> | | | |
| $XY_{min}$ | <0.0,0.1> | | | |
| Ref | <0.4,0.6> | | | |
| | | 0.2236✗ | 0 | 0.2236✗ |
| Actual distance w.r.t reference STS = 0.32 | | | | |

Table 8. Detailed LB distance of pattern [XY]

| Itemset | STS | Upper-lower & Lower-lower bounds w.r.t reference | | |
|---------|-----|------|------|------|
| | | ULB | LLB | LB |
| X | <0.6,0.4> | | | |
| $\overline{Y}$ | <0.7,0.3> | | | |
| $X\overline{Y}_{UBSTS}$ | <0.6,0.3> | | | |
| $X\overline{Y}_{LBSTS}$ | <0.3,0.0> | | | |
| $XY_{MIN}$ | <0.0,0.1> | | | |
| $XY_{MAX}$ | <0.3,0.4> | | | |
| Ref | <0.4,0.6> | | | |
| | | 0.2236 | 0.0 | 0.2236 |
| Similarity of temporal pattern [XY] w.r.t Ref | | | | ✗ |
| Move to Level-3 | | | | ✗ |

Table 10. Detailed LB distance of itemset [XZ]

| Itemset | STS | Upper-lower & Lower-lower bounds w.r.t reference | | |
|---------|-----|------|------|------|
| | | ULB | LLB | LB |
| X | <0.6,0.4> | | | |
| $\overline{Z}$ | <0.2,0.2> | | | |
| $X\overline{Z}_{UBSTS}$ | <0.2,0.2> | | | |
| $X\overline{Z}_{LBSTS}$ | <0.0,0.0> | | | |
| $XZ_{MIN}$ | <0.4,0.2> | | | |
| $XZ_{MAX}$ | <0.6,0.4> | | | |
| Ref | <0.4,0.6> | | | |
| | | 0.2 | 0.0 | 0.2 |
| Similarity of temporal pattern [XZ] w.r.t Ref | | | | ✓ |
| Move to Level-3 | | | | ✓ |

Table 9. Lower Bound distance computation of XZ

| Itemset | STS | ULB | LLB | LB |
|---------|-----|-----|-----|-----|
| $XZ_{max}$ | <0.6,0.4> | | | |
| $XZ_{min}$ | <0.4,0.2> | | | |
| Ref | <0.4,0.6> | | | |
| | | 0.2✓ | 0 | 0.2✓ |
| Actual distance w.r.t reference STS = 0. 2 | | | | |

Table 12. Detailed LB distance of itemset [YZ]

| Itemset | STS | Upper-lower & Lower-lower bounds w.r.t reference | | |
|---------|-----|------|------|------|
| | | ULB | LLB | LB |
| Y | <0.3,0.7> | | | |
| $\overline{Z}$ | <0.2,0.2> | | | |
| $Y\overline{Z}_{UBSTS}$ | <0.2,0.2> | | | |
| $Y\overline{Z}_{LBSTS}$ | <0.0,0.0> | | | |
| $YZ_{MIN}$ | <0.1,0.5> | | | |
| $YZ_{MAX}$ | <0.3,0.7> | | | |
| Ref | <0.4,0.6> | | | |
| | | 0.1 | 0.0 | 0.1 |
| Similarity of temporal pattern [YZ] w.r.t Ref | | | ✓ | |
| Move to Level-3 | | | | ✓ |

Table 11. Lower Bound distance computation of YZ

| Itemset | STS | ULB | LLB | LB |
|---------|-----|-----|-----|-----|
| $YZ_{max}$ | <0.3,0.7> | | | |
| $YZ_{min}$ | <0.1,0.5> | | | |
| Ref | <0.4,0.6> | | | |
| | | 0.1✓ | 0 | 0.1✓ |
| Actual distance w.r.t reference STS = 0. 1414 | | | | |

**Step2:**

Compute Positive and Negative support time sequences $(S_{\theta_P}, S_{\theta_N})$ of singleton temporal items from support values of corresponding positive and negative items obtained in step-1. We can obtain the positive and negative support sequences of singleton items as shown in Table.5 for time slots $t_1$ and $t_2$ from Table.4.

**Step-3:**

Find Level -1 similarity profiled temporal items. In this step, we compute both the approximate upper lower bound (ULB) and actual Euclidean distance between reference vector and singleton items. If the actual distance is less than the user threshold, this means that the singleton item is temporally similar. Otherwise it is temporally dissimilar pattern.

We also find the approximate upper lower bound distance value which is the deciding factor to consider or discard the corresponding singleton item. In the present example, even though true distances of X and Z do not satisfy the threshold constraint, we consider retaining these singleton items, as these may be helpful to generate level-2 itemsets. We retain all such items whose upper-lower bound distance satisfies the threshold constraint as shown in Table.6.

**Step-4: Find Temporal Patterns of Size, |S|=2**

This step involves generating support sequence for item sets of length, |S|=2, as denoted by [XY], [XZ] and [YZ]. From these support sequences generated, we must find whether itemsets satisfy threshold constraint. This must be done without scanning the database. We now show computations of these itemsets.

**A. Finding Support Time Sequence for temporal pattern [XY]**

The figure.1 shows the Venn diagram for pattern [XY]. Here X' and Y' indicates complementary or negative pattern.

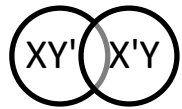

Figure.1 Computation of Support sequence for XY using Venn diagram

Here $I_i = X$ and $I_j = Y$. The temporal support sequence for itemset XY may be obtained, either using expression

$$XY = X - X\overline{Y} \quad (Or)$$

$$XY = Y - \overline{X}Y \qquad \text{[Note: both methods give same values, left to reader to evaluate]}$$

Now, using procedure to find upper and lower bound support sequences as discussed in section 3.7, we have

$$X\overline{Y}_{UBSTS} = < \min(0.6, 0.7), \min(0.4, 0.3) > = < 0.6, 0.3 >$$

$$X\overline{Y}_{LBSTS} = < \max(0.6 + 0.7 - 1, 0), \max(0.4 + 0.3 - 1, 0) > = < 0.3, 0.0 >$$

$$XY = \begin{cases} X - X\overline{Y}_{UBSTS} = < 0.0, 0.1 > = XY_{MIN} \\ X - X\overline{Y}_{LBSTS} = < 0.3, 0.4 > = XY_{MAX} \end{cases}$$

From these minimum and maximum support sequences obtained for the temporal pattern [$X\overline{Y}$], we can deduce maximum and minimum possible support sequences for temporal pattern [XY] defined as, $XY_{min} = < 0.0, 0.1 >$ and $XY_{max} = < 0.3, 0.4 >$. Table.7 gives computation of lower bound distance of temporal pattern [XY]. Since, the upper lower bound and lower bound distances of support sequence of temporal itemset [XY] w.r.t reference support sequence, R does not satisfy user defined threshold value constraint; we consider the temporal itemset [XY] as temporally dissimilar pattern w.r.t R. The Table.8 shows detailed computation process of support sequence, support values for temporal pattern [XY].

**B. Finding Support Time Sequence for temporal pattern [XZ]**

The Figure .2 shows the Venn diagram for pattern [XZ]. Here X' and Z' indicates complementary pattern.

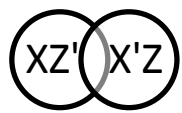

Figure.2 Computation of Support sequence for XZ using Venn diagram

Here $I_i = X$ and $I_j = Z$. We may obtain support sequence for XZ using two ways, either using the expression

$$XZ = X - X\bar{Z} \quad (or)$$

$$XZ = Z - \bar{X}Z$$

Now, using procedure to find upper and lower bound support sequences as discussed above we have,

$$X\bar{Z}_{UBSTS} = \; < \min(0.6, 0.2), \min(0.4, 0.2) > \; = \; < 0.2, 0.2 >$$

$$X\bar{Z}_{LBSTS} = \; < \max(0.6 + 0.2 - 1, 0), \max(0.4 + 0.2 - 1, 0) > \; = \; < 0, 0 >$$

$$XZ = \begin{cases} X - XZ_{UBSTS} \; = \; < 0.4, 0.2 > \; = \; XZ_{MIN} \\ X - XZ_{LBSTS} \; = \; < 0.6, 0.4 > \; = \; XZ_{MAX} \end{cases}$$

From these lower bound and upper bound support sequences obtained for the temporal pattern $[X\bar{Z}]$, we can deduce maximum and minimum possible support sequences for temporal pattern $[XZ]$ defined as, $XZ_{min} = \; < 0.4, 0.2 >$ and $XZ_{max} = \; < 0.6, 0.4 >$. Table.9 gives computation of lower bound distance of temporal pattern [XZ]. Since, the upper lower bound and lower bound distances of support sequence of temporal itemset [XZ] w.r.t reference support sequence, R satisfies user defined threshold value constraint; we consider the temporal itemset [XZ] as temporally similar pattern w.r.t R. The table 10 shows the detailed computation process of support sequence, support values for temporal pattern [XZ].

### C. Finding Support Time Sequence for temporal pattern [YZ]

Here $I_i = Y$ and $I_j = Z$. We may obtain support sequence for temporal pattern [YZ] using two ways, either using the expression

$$YZ = Y - Y\bar{Z} \quad or$$

$$YZ = Z - \bar{Y}Z$$

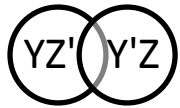

Figure.3 Computation of Support sequence for YZ using Venn diagram

The Figure.3 shows the Venn diagram for pattern [YZ]. Here Z' and Y' indicates complementary or negative pattern. Now, using procedure to find upper and lower bound support sequences as discussed above we have,

$$Y\bar{Z}_{UBSTS} = \; < \min(0.3, 0.2), \min(0.7, 0.2) > \; = \; < 0.2, 0.2 >$$

$$Y\bar{Z}_{LBSTS} = \; < \max(0.3 + 0.2 - 1, 0), \; \max(0.7 + 0.2 - 1, 0) > \; = \; < 0, 0 >$$

$$YZ = \begin{cases} Y - Y\bar{Z}_{UBSTS} \; = \; < 0.1, 0.5 > \; = \; YZ_{MIN} \\ Y - Y\bar{Z}_{LBSTS} \; = \; < 0.3, 0.7 > \; = \; YZ_{MAX} \end{cases}$$

From these lower bound and upper bound support sequences obtained for the temporal pattern $[Y\bar{Z}]$, we can deduce maximum and minimum possible support sequences for temporal pattern [YZ] defined as, $YZ_{min} = \; < 0.1, 0.5 >$ and $YZ_{max} = \; < 03, 0.7 >$. Table.11 gives computation of lower bound distance of temporal pattern [YZ]. Since, the upper lower bound and lower bound distances of support sequence of temporal itemset [YZ] w.r.t reference support sequence, R satisfies user defined threshold value constraint; we consider the temporal itemset [YZ] as temporally similar pattern w.r.t R. Table.12 shows detailed process of computation of support sequence, support values for the temporal association pattern [YZ].

**Step-5: Generate Temporal Pattern of Size, |S|=3, XYZ**

Consider upper lower bound distance and lower bound distance of support sequences of association patterns [XY], [XZ] and [YZ] w.r.t reference sequence R. Here ✓ and ✗ indicates satisfying and not satisfying the threshold constraint respectively. Since, only [XZ] and [YZ] satisfy threshold constraint, and [XY] is not similar w.r.t the reference sequence, we consider association pattern [XYZ] to be temporally dissimilar. This is because subsets of the pattern [XYZ] which are [XY], [XZ], and [YZ] must also be temporally similar which is not true w.r.t pattern [XY]. Thus, final set of temporal association patterns which are similar to reference are [Y], [XZ] and [YZ].

Alternately, we may compute lower bound distance of [XYZ] by using the Equation.1 and verify, if the association pattern [XYZ] is temporally similar or not. This process is explained below.

Here $I_i = XY$ and $I_j = Z$. We may obtain support sequence for temporal pattern [XYZ] as explained below

We give equation.1 again as shown below,

$$I_i\,I_j = \begin{cases} I_{i_{min}} - I_{i_{min}} \otimes \overline{I_j} \\ I_{i_{max}} - I_{i_{max}} \otimes \overline{I_j} \end{cases}$$

Here $I_i = XY$ and $I_j = Z$. We can find support sequence for $I_i\,I_j$ using Equation.1 as shown below

$$XYZ = \begin{cases} XY_{min} - XY_{min} \otimes \overline{I_j} \\ XY_{max} - XY_{max} \otimes \overline{I_j} \end{cases}$$

We map $I_{i_{max}}$, $I_{i_{min}}$ and $\overline{I_j}$ to $XY_{MAX}$, $XY_{MIN}$ and $\bar{Z}$ respectively as shown below

$I_{i_{max}} = XY_{MAX} = <0.3, 0.4>$
$I_{i_{min}} = XY_{MIN} = <0.0, 0.1>$
$\overline{I_j} = \bar{Z} = <0.2, 0.2>$

So, the support sequence for [XYZ] may be obtained using computations as shown below

$[XYZ] =$
$$= \begin{cases} <0.0,0.1> - <0.0,0.1> \otimes <0.2,0.2> \\ <0.3,0.4> - <0.3,0.4> \otimes <0.2,0.2> \end{cases}$$

$$= \begin{cases} <0.0,0.1> - <0.0,0.1> \\ <0.0,0.1> - <0.0,0.0> \\ <0.3,0.4> - <0.2,0.2> \\ <0.3,0.4> - <0.0,0.0> \end{cases}$$

$$= \begin{cases} <0.0,0.0> \\ <0.0,0.1> \\ <0.1,0.2> \\ <0.3,0.4> \end{cases}$$

From these support sequences obtained, we can obtain maximum and minimum possible support sequences for XYZ as

$[XYZ]_{max} = <0.3, 0.4>$ and
$[XYZ]_{min} = <0.0, 0.0>$

Table 13. Lower Bound distance computation of XYZ

| Itemset | STS | ULB | LLB | LB |
|---|---|---|---|---|
| $XYZ_{max}$ | <0.3,0.4> | | | |
| $XYZ_{min}$ | <0.0,0.0> | | | |
| Ref | <0.4,0.6> | | | |
| | | 0.2236× | 0 | 0.2236× |
| Actual distance w.r.t reference STS = 0. 3162 × | | | | |

From Table.13, we can see that the lower bound distance is 0.2236 and exceeds threshold value. Hence, the itemset [XYZ] is not temporally similar w.r.t reference.

## 5. CONCLUSIONS

In this research, given a reference support time sequence defined for a finite number of time slots, we aim to discover all such temporal association patterns from temporal database which are similar to the reference vector. We introduce the concept of negative itemset support values and negative itemset support sequences. In essence, we maintain two types of support values, called positive and negative support values. In addition, we also maintain two types of support sequences, called positive and negative support sequences. The idea is to use the concept of Venn diagrams, to find similar temporal association patterns satisfying given threshold value, with primary objective of eliminating multiple temporal database scans. This is achieved by reuse of support sequences of temporal sub patterns computed in previous stages. The proposed approach thus reduces the space and time complexities. In future, we there is a scope for extending present approach by designing a novel similarity measure to find similar temporal patterns.